\documentstyle[preprint,aps,epsfig,12pt]{revtex}
\def\nuc#1#2{\mbox{${}^{#1}$\hbox{#2}}}
\def\Au{\nuc{197}{Au}}
\def\Pb{\nuc{208}{Pb}}
\def\AGeV{\mbox{$A$\hbox{~GeV}}}
\begin{document}
\thispagestyle{empty}
\draft
\tighten
\titlepage
\title{
	Re-Hardening of Hadron Transverse Mass Spectra
	in Relativistic Heavy-Ion Collisions
}

\author{N. Otuka$^{\rm ab}$, 
	P.K. Sahu$^{\rm a}$, 
	M. Isse$^{\rm a}$, 
	Y. Nara$^{\rm c}$
	and
	A. Ohnishi$^{\rm a}$
	}

\address{%
	a\ Division of Physics, Graduate School of Science,
                Hokkaido University, Sapporo 060-0810, Japan\\
	b\ Advanced Science Research Center,
		Japan Atomic Energy Research Institute,\\
		Tokai, Ibaraki 319-11, 
		Japan\\
	c\ Physics Department, Brookhaven National Laboratory,
         Upton, N.Y. 11973, U.S.A.}

\date{\today}

\maketitle

\begin{abstract}
We analyze the spectra of pions and protons
in heavy-ion collisions at relativistic energies
from 2 \AGeV to 65+65 \AGeV
by using a jet-implemented hadron-string cascade model.
%pks , JAM.
%
In this energy region,
hadron transverse mass spectra first show softening
until SPS energies, and re-hardening may emerge at RHIC energies.
Since hadronic matter is expected to show only softening
at higher energy densities, this re-hardening of spectra
can be 
%pks
interpreted as 
%pks
a good signature of the quark-gluon plasma formation.
\end{abstract}

\bigskip

\endtitlepage

%%%%%%%%%%%%%
%\section{Introduction}
%%%%%%%%%%%%%

%%%%%%%%%%%%%
% General Introduction
%%%%%%%%%%%%%
The 
%pks largest aim 
main goal %pks or greatest interest
%pks
of the high-energy heavy-ion collisions is to
explore hot and/or dense hadronic matter
far from stable nuclei.
Especially, the formation of the quark-gluon plasma (QGP)
is of the primary interest.
In order to achieve this, %pks QGP formation, 
%pks significant
sincere
%pks 
efforts have been %pks devoted 
made 
%pks
in this decade,
starting from Bevalac at LBL,
followed by 
GSI-SIS, BNL-AGS, CERN-SPS, and BNL-RHIC,
by increasing incident energies~\cite{QM1997,QM1999}.
%
%pks Up to SPS energies,
The QGP is expected to be formed at SPS energies, but
%pks
we have not yet verified 
%pks the QGP formation
this 
%pks
clearly,
although there are some promising signals;
the anomalous charmonium suppression~\cite{Matsui1986},
enhanced production of strange hadrons~\cite{Rafelski1991},
and enhanced low-mass di-lepton production~\cite{GQLi1995},
suggest the formation of anomalous hadronic matter
in which the chiral symmetry is partially restored 
and non-negligible color-flux is stored.
However, we have not %pks found
seen 
%pks
any evidence which clearly suggests
the bulk formation of QGP.

%%%%%%%%%%%%%
% Softening
%%%%%%%%%%%%%
The most characteristic feature of the QGP formation 
is the sudden release of a large number of degrees of freedom (DOFs),
and the consequent softening at around the critical temperature,
but this softening can be mimicked 
%pks also 
by hadronic (resonance and string) DOFs.
%
%pks Actually,
Practically,
%pks
the softening of matter is observed and recognized
already at SIS-AGS-SPS energies,
where the inverse slope parameters of hadron spectra 
seem to be saturated,
and radial and directed flows are decreasing~\cite{Herrmann1996}.
For example, the directed flow exhibits a maximum at SIS energies,
and goes down at higher energies.
Since QGP formation is not expected at SIS energies,
this softening should be caused by hadronic origin.
In Ref.~\cite{SCMO2000}, Sahu et al. have shown that the incident energy
dependence of directed and elliptic flows as well as 
transverse mass spectrum of protons are well 
%pks explained
reproduced %pks or described
%pks
within a hadron-string picture,
by taking into account the reduction of repulsive nuclear interaction
and the appropriate increase of resonance-string DOFs.
Hagedorn discussed the role of the latter in 1965~\cite{Hagedorn1965},
suggesting that a limiting temperature around the pion mass 
would appear if the hadronic level density
grows exponentially as a function of mass.
In this case, a large part of energy is exhausted by the mass energies of
heavy-hadrons, then the kinetic energy per hadron cannot be larger than 
some 
%pks
particular %pks What is the value???
%pks
value. This also implies that the pressure would not grow as rapidly as
the energy density, 
%pks namely 
as a result
%pks
the softening of hadronic matter occurs.
Thus the above softening can be considered as a partial realization 
of Hagedorn gas, where large hadronic DOFs are activated.
In a more dynamical context, 
many of the hadronic cascade 
models~\cite{Sorge1989,Cassing1990,Amelin1990,Bass1998,NOONC2000}
which explain the data at these energy region
incorporate
large hadronic DOFs, including various resonances and strings,
then they naturally describe the above  softening.
Although there are several models~\cite{Pang1992,BALi1995} 
which contain smaller hadronic DOFs and explain the data, 
they usually incorporate multi-particle production 
with a finite formation time,
which plays a role to generate effective large DOFs~\cite{OtukaPhD}.

%%%%%%%%%%%%%
% Re-Hardening
%%%%%%%%%%%%%
The above discussion tells us that 
%pks we cannot 
it is very hard to  % or it is difficult to
%pks
verify the QGP formation
only from the softening of matter,
since we cannot distinguish the effects of QGP from those due to
the increase of hadronic DOFs.
On the other hand, if a re-hardening is observed at higher energy densities,
it is very hard to explain in a hadronic scenario.
%
%pks Actually, recently obtained data at 
The data obtained very recently from
%pks 
RHIC energy seem 
to exhibit this signature~\cite{HOhnishi2000,NuXu2001}. 
%pks ,HOhnishi2000}.
%
The estimated radial flow velocity at $\sqrt{s}$ =130 \AGeV
is much larger than that at SPS energies.
Since the radial flow velocity grows until around AGS energies
and stays constant or decreases 
(depending on the momentum range to estimate the inverse slope parameter)
between AGS and SPS for heavy systems, 
the above increase suggests that the system becomes hard again at RHIC.

%%%%%%%%%%%%%
% Re-Hardening
%%%%%%%%%%%%%
The %pks above 
idea of hardening after reaching QGP is not very new.
For example, 
a steep increase of an average transverse momentum was observed
at around an energy density of 1.5 GeV/fm$^3$
in the high energy cosmic ray nuclear interactions
on emulsion chambers by JACEE collaboration~\cite{JACEE1987}.
From the theoretical side, 
Bass et al. have recently shown that a similar tendency is
expected at RHIC and LHC energies
by applying a hydrodynamical model with UrQMD after burner~\cite{Bass2000}.
The origin of this increase of transverse momentum
is nothing but the steep increase of the pressure in QGP.
%
%Recently these softening and re-hardening are revealed
%by applying hydrodynamical model and microscopic nonequilibrium
%model at deconfined stage and hadronization stage
%to the SPS, RHIC and LHC energies~\cite{Bass2000}. 
%
%However, partons are not ingredients in this model.
%%
However, in this model, 
%pks treatment,
the role of initial parton dynamics such as mini-jet production is 
not taken into account,
and the local equilibrium is assumed {\it a priori}
in the first stage of heavy-ion collisions.
\medskip

%%%%%%%%%%%
% Purpose
%%%%%%%%%%%
Thus at present, it is very interesting and urgently desired
to examine the %pks above 
behavior of radial flow
by using models having both of hadronic and partonic aspects.
In this 
%pks paper,
letter,
%pks 
we show 
%pks the calculated results of 
%pks
the transverse mass spectra of hadrons
from SIS to RHIC energies, by using a jet-implemented hadron-string 
cascade model, JAM~\cite{NOONC2000}.

%%%%%%%%%%%
%\section{Model}
%%%%%%%%%%%

In JAM, various hadronic resonances as well as strings are  
explicitly propagated.
At high energies ($\sqrt{s}>$10 GeV),
we also include multiple mini-jet production 
in which jet cross section and the jet number
are calculated using an eikonal formalism for
perturbative QCD and hard parton-parton scattering
with initial and final state radiation are simulated
using the Lund model (PYTHIA 6.1)~\cite{Sjostrand1994}.
In this framework,
we can simulate parton and gluon DOFs belong to one
parton-parton hard scattering.
In addition, it has been already demonstrated that 
JAM explains the hadron spectra very well 
from $p$+\nuc{9}{Be} to \nuc{197}{Au}+\nuc{197}{Au} reactions
at AGS energies~\cite{NOONC2000},
and those in \nuc{208}{Pb}+\nuc{208}{Pb} reactions
at SPS energies~\cite{Nara1998}.
At collider energies, JAM can be considered as a space-time version of
HIJING~\cite{Wang1994,Wang1997}.
Therefore, it is an appropriate framework to describe the bulk behavior
of hadron spectra in a wide energy 
%pks region 
range,
%pks
systematically.

\medskip

%%%%%%%%%%%
%\section{Results}
%%%%%%%%%%%
We have made simulation calculations
of \Au+\Au\ reactions
	at SIS (2 \AGeV), AGS (10.6 \AGeV), JHF (25 \AGeV)
	and RHIC ($\sqrt{s}$ = 56 and 130 \AGeV)
	energies,
and \Pb+\Pb\ reactions
	at SPS (158 \AGeV) energy.
In these calculations,
the open source program, JAM1.0
has been used with default parameters~\footnote{
We have adopted the source code of JAM version 1.009.27 (April 21, 2000),
which was written before the RHIC experiments started,
to avoid any fitting to the data.
}.
In all of the reactions,
the impact parameter range is limited to 
$0 < b < 3.3$ fm, which corresponds to central 350 mb collisions.
In each incident energy,
we have generated more than 1000 events.
%which is enough to verify the trend.

\begin{figure}
\epsfig{file=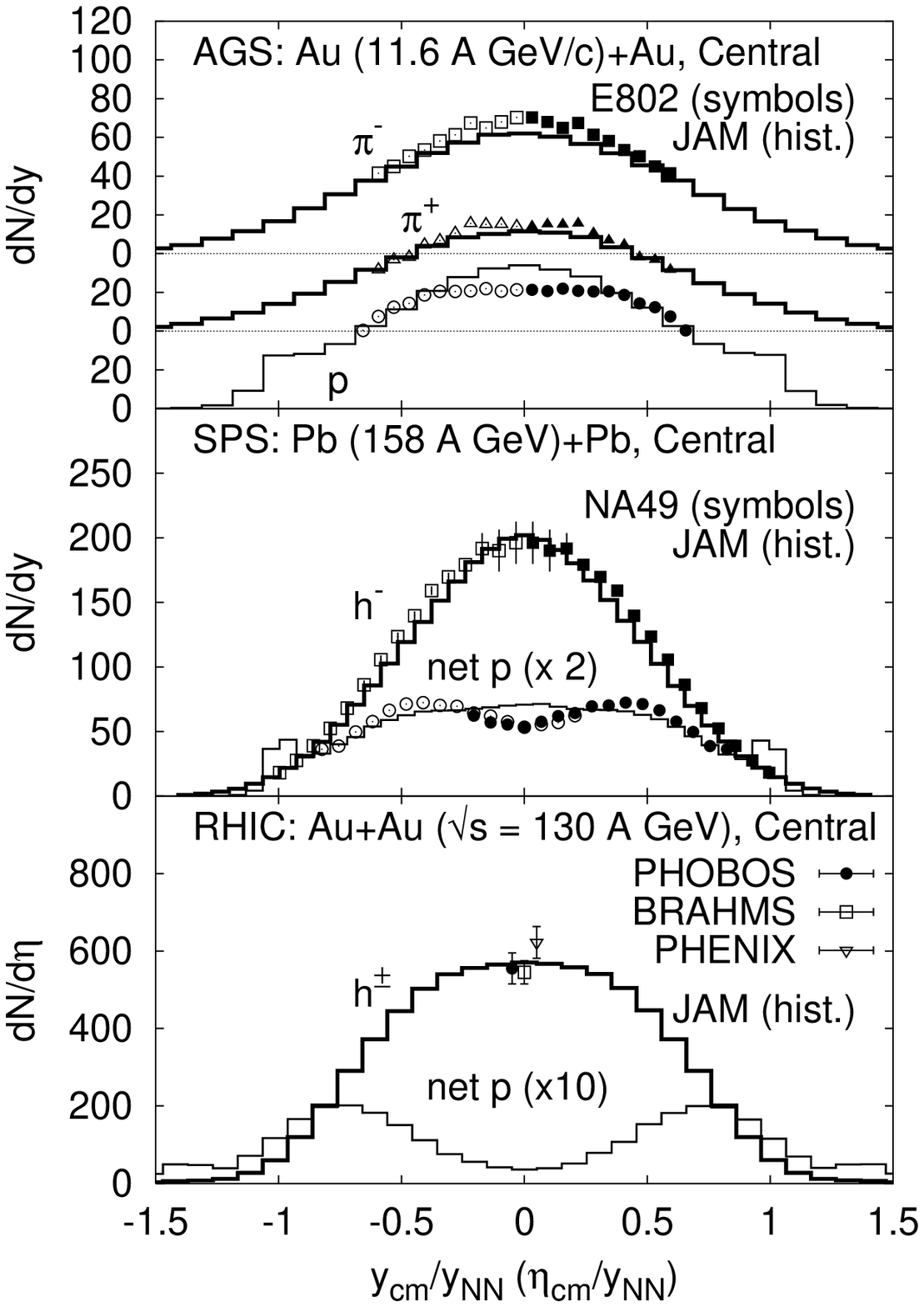,width=14cm}%
\caption{
Rapidity distribution $(dN/dy)$ at the AGS and SPS energies,
and 
pseudorapidity distribution $(dN/d\eta)$ at the RHIC energy.
Calculated results are compared with the 
E802~\protect\cite{E802},
NA49~\protect\cite{NA49},
PHENIX~\protect\cite{PHENIX2000},
BRAHMS~\protect\cite{BRAHMS2001},
and
PHOBOS~\protect\cite{PHOBOS2000}
data.
Collisions with impact parameter $b<$3.3 fm has been taken
in the calculations.
For experiments,
$\sigma_{\rm trig}$=350 mb for the E802 experiment,
5\% for NA49 and PHENIX,
and 6\% for BRAHMS and PHOBOS.
}
\label{fig:dNdY}
\end{figure}

%%%%%%%%%%%
% dN/dY
%%%%%%%%%%%
In Fig.~\ref{fig:dNdY},
we show the rapidity distributions of hadrons
at AGS, SPS and RHIC energies.
Although the baryon stopping power is overestimated a little at AGS and SPS
energies, the 
%pks
over all
%pks
general trend of data is well reproduced.
At RHIC energy ($\sqrt{s}$ = 130 \AGeV), 
there are three sets of data from 
PHENIX collaboration~\cite{PHENIX2000},
BRAHMS collaboration~\cite{BRAHMS2001},
and
PHOBOS collaboration~\cite{PHOBOS2000}, 
%pks.
for the charged particle pseudo-rapidity distributions at mid-rapidity
%pks
region.
%pks
%pks There are three data why both?? 
%pks Both of the data 
All of them
%pks
give the value around 
$dN/d\eta|_{\eta=0} \simeq 570$ for central collisions,
which is well reproduced, as well.
In addition, the calculated
ratio of pseudorapidity density of ${\bar p}$ to $p$ at $\eta=0$ is 0.63
which agrees well with experimental data of
0.61$\pm$0.06(stat.)$\pm$0.4(syst.) from 
the BRAHMS collaboration~\cite{BRAHMS2001}.
Therefore, we can expect that the description of the bulk dynamics
from AGS to RHIC energies within JAM is reliable.

\begin{figure}
\epsfig{file=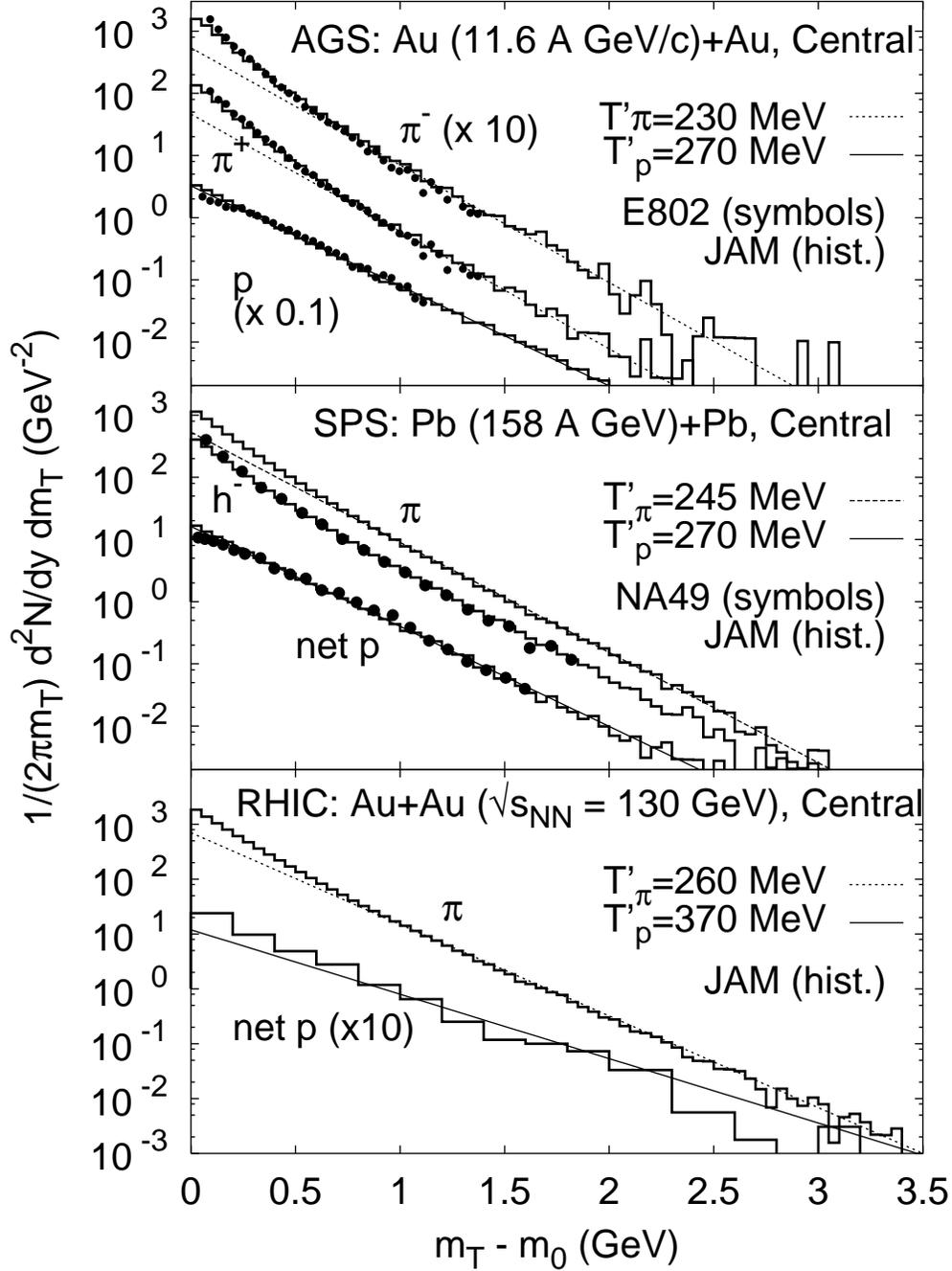}%
\caption{
Transverse mass spectra of hadrons at the AGS, SPS and RHIC energies.
Calculated results are compared with the
E802~\protect\cite{E802}
and 
NA49~\protect\cite{NA49} data.
Collisions with impact parameter $b<$3.3 fm has been taken
in the calculations.
For experiments,
$\sigma_{\rm trig}$=350 mb for the E802 experiment,
and 5\% for NA49.
The exponential lines with the slope parameters $T'$
are shown to guide eyes.
}
\label{fig:dMT}
\end{figure}

%%%%%%%%%%%
% d^2N/Mt/dMt/dY
%%%%%%%%%%%
Next we show the transverse mass spectra of hadrons at mid-rapidity
in Fig.~\ref{fig:dMT}.
At AGS and SPS energies,
the pion and negative hadron spectra are very well reproduced,
including the deviation from a single exponential behavior
at low energies,
coming from the decay of low-lying baryon resonances.
For protons,
the higher energy behavior is satisfactory, 
but the lower energy part is overestimated.
Since the yield of this part is known to be very sensitive
to the nuclear mean field~\cite{SCMO2000}, 
this overestimate would be a natural consequence of cascade model results
without mean field.

%It is very interesting to find that the 
In this figure, we find an interesting behavior of proton spectra.
The inverse slope parameters of protons at AGS and SPS energies are  
almost the same, but grows rapidly at RHIC energy.
On the other hand, pion spectra become stiffer gradually 
as the incident energy increases.
%

\begin{figure}
\epsfig{file=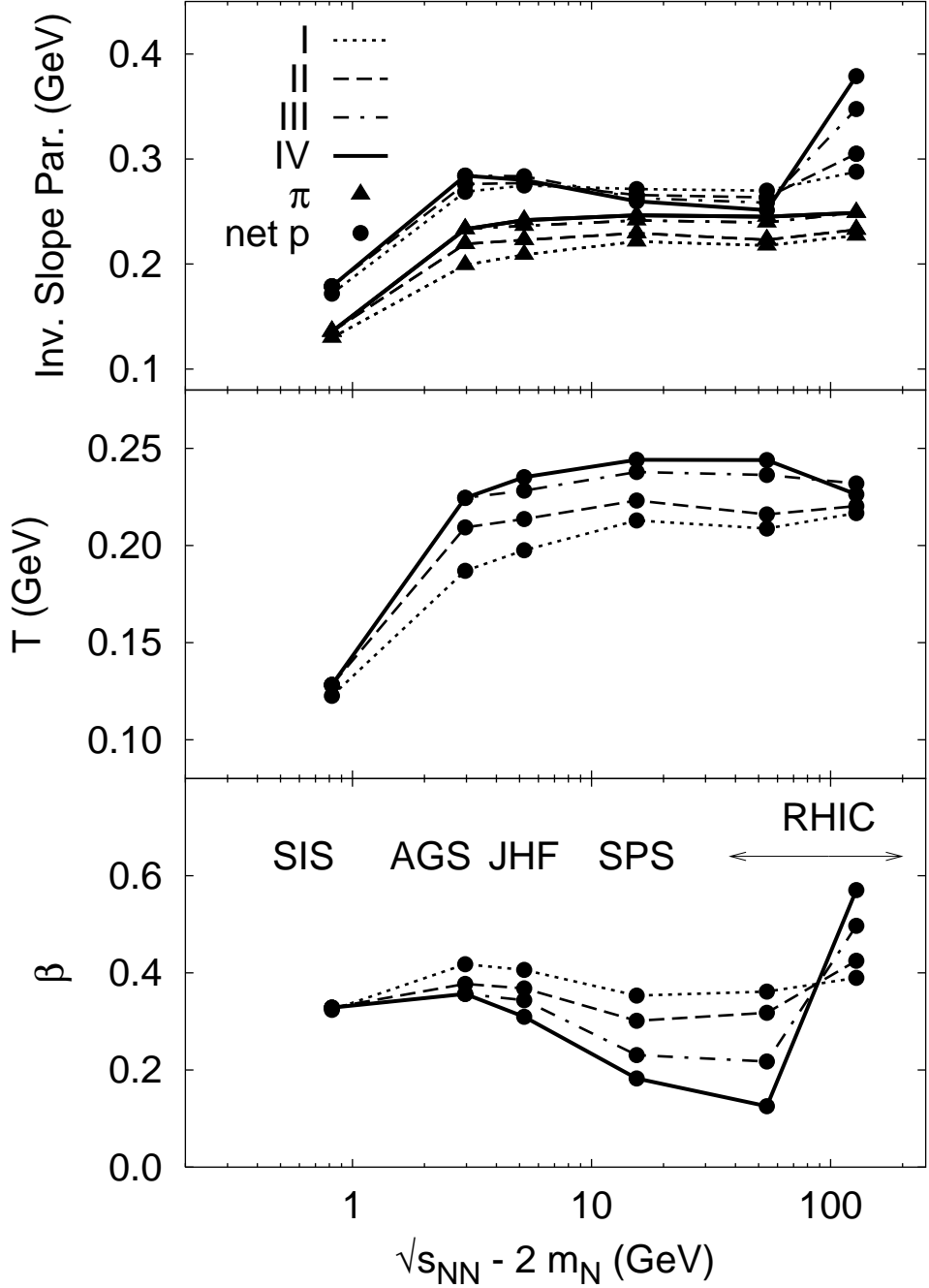}%
\caption{
Calculated inverse slope parameters (top)
and the extracted temperature (middle) 
and radial flow parameters (bottom) from SIS to RHIC energies
with three fit ranges sets of I (dotted), II (dashed), III (dot-dashed)
 and IV (solid).
See Table \protect\ref{tab:Mmin} for fit ranges I, II, III and IV.
}
\label{fig:Slope}
\end{figure}

%%%%%%%%%%%
% T and beta
%%%%%%%%%%%
In order to 
%pks discuss 
realize
%pks
this point more quantitatively, 
we fit the transverse mass spectra of pions and protons
with a single exponential.
As described above, the low energy part of the spectra
is affected by other mechanism than the emission from 
an expanding fire ball, 
therefore we chose the energy region, 
$\Delta_{\rm min}<m_T - m_0 < \Delta_{\rm max}$.
For the choice of $\Delta$, we have tried several cases,
as shown in Table~\ref{tab:Mmin}.
The results of fitting are shown in the upper panel of Fig.~\ref{fig:Slope}.
In all the cases,
the inverse slope parameter of proton stays almost constant 
between AGS and RHIC energies at $\sqrt{s}$ = 56 \AGeV,
and rapidly grows at RHIC energy of $\sqrt{s}$ = 130 \AGeV.
When we separate the inverse slope parameter ($T'$)
into the temperature ($T$) and radial flow ($\beta_t$)
by using the non-relativistic relation,
$T' = T + m \beta_t^2/2$,
the meaning of the above behavior becomes clearer.
As shown in the middle and the lower panel of Fig.~\ref{fig:Slope}, 
while the hadronic temperature slowly grows as a function of incident energy,
the flow velocity first grows at low energies, subsequently saturates,
then decreases between AGS and SPS energies,
and increases {\em again} at RHIC energies.

%%%%%%%%%%%
% Interpretation
%%%%%%%%%%%
This behavior --- re-hardening after softening --- 
can be most naturally interpreted with a phase transition scenario
from hadronic matter at the Hagedorn regime to QGP.
Hadronic matter becomes softer according to the large level density of
hadronic objects, but at some energy density, 
hadrons cannot 
%pks be
exist
%pks
as they are in vacuum to dissolve into quarks and gluons.
Then the pressure grows linearly again as a function of the energy density,
$dP/d\epsilon \simeq 1/3$.
On the other hand, 
it would be very difficult to interpret the above behavior
in purely hadronic scenarios.
We have to assume a very strong repulsive interaction at very high energy
density, or we have to assume a rapid decrease of hadronic level density
at some mass.
As for the former, it is already shown that the reduction of repulsive
interaction at high momenta or density is necessary in the analysis 
of directed and elliptic flows at SIS to AGS energies~\cite{SCMO2000},
then it is unnatural to incorporate very strong repulsive interaction 
again at RHIC energies.
The latter is also an unnatural assumption,
because of the complex particle nature of hadrons.

As shown, the behavior of temperature and radial flow depends
on the selection of the fitting range.
The lower part of transverse spectra is dominated by 
resonance decay at later stage of collisions,
while mini-jet production affects on higher part.
It means transverse mass spectra are involved
by various stage of collisions.
So the more detail study on the relation between
the fitting range and collision history is needed.

\medskip

%%%%%%%%%%%
%\section{Discussions and Conclusion}
%%%%%%%%%%%
In this 
%pks paper,
letter, 
%pks
we have demonstrated that
the re-hardening of hadron spectra would emerge at RHIC energies.
%even if we consider hadronic interaction.
%
The most natural interpretation for this re-hardening 
is achieved by assuming the QGP formation,
and this behavior is already seen in the preliminary data
at RHIC~\cite{NuXu2001}.
%
%Preliminary data at RHIC energies from the PHENIX collaboration~\cite{H_Ohnishi}
%support this re-hardening.
%
Confirmation and further studies are necessary
%pks
to conclude firmly.
%pks

%%%%%%%%%%%%%%%%%%%%%%
% To Yasushi Nara: Read here.
%%%%%%%%%%%%%%%%%%%%%%
The above interpretation is based on a hydrodynamical picture
of heavy-ion collisions, 
and this is supported by the recent RHIC data,
which suggest that equilibrium is reached to a certain rate.
The hadronic transverse mass spectra shows exponential behavior
rather than a power law behavior, 
and the inverse slope parameter behaves linearly 
as a function of mass, approximately~\cite{QM2001}.
These facts are consistent with a hydrodynamical picture
of the expanding fireball.
Equilibration processes largely affect the hadron spectra
in the present calculation, too.
For example, if we ignore meson-baryon and meson-meson collisions in JAM, 
hadron spectra at RHIC become much softer at low momenta
and strongly deviate from 
%pks a 
the
%pks
exponential behavior at large momenta.
Therefore, the hadron interactions in the later stage are very important,
and the hadron gas is well equilibrated.
On the other hand, the partonic equilibration among different mini-jets 
(parton cascade) is not included in the present model treatment.
However, since the bulk part of hadrons are strongly kicked by 
initial mini-jets, the space-time volume where partons are propagating
is considered to be large.
Then it is natural to expect that equilibration among mini-jets to QGP
proceeds easily, once the parton cascade processes are incorporated.
These parton cascade processes would modify the present results
in a better 
%pks 
constructive 
%pks
direction.
For example, the preliminary RHIC data 
shows stiffer hadron transverse mass spectra
and a larger baryon stopping power ($dN(\hbox{net }p)/dy \geq 10$)
than the calculated results shown here.
%%%%%%%%%%%%%%%%%%%%%%

% Byproduct.
We have also demonstrated that the recently developed
jet-implemented hadron-string cascade model, JAM, is capable of describing
the bulk dynamics of high-energy heavy-ion collisions from AGS to RHIC
energies.
Within this model with default parameters,
the local minimum of the radial flow is calculated to appear 
between SPS and lower RHIC energies,
and the local maximum of radial flow in the hadronic regime
between AGS and JHF energies.
Therefore, the excitation function between SPS and RHIC energies should be
useful to get the signature of the quark-gluon plasma formation.

\medskip

We would like to thank Dr. H. Ohnishi, Professor R.S. Hayano, 
and Professor S. Muroya for useful discussions and suggestions.
This work was supported in part by
the Grant-in-Aid for Scientific Research
(No.\ 09640329) % Kiban C (Fragmentation, A.O.) 1997-2000
from the Ministry of Education, Science and Culture, Japan.
The calculations were partially supported by
Hierarchical Matter Analyzing System 
at the Division of Physics, Graduate School of Science,
Hokkaido university.
One of the authors (P.K.S.) acknowledges the support 
of the Japan Society for the Promotion of Science (ID No. P98357).

%%%%%%%%%%%%%%%%%%%%%%%%%%%%%%%%%%%%%%%%%%%%%%%%%%%%%%%%%%%%%%%%%%%%%%%%%%%%%%%
%                       REFERENCES:
%%%%%%%%%%%%%%%%%%%%%%%%%%%%%%%%%%%%%%%%%%%%%%%%%%%%%%%%%%%%%%%%%%%%%%%%%%%%%%%

\newpage
%%%%%%%%%%%%%%%%%%%%%%%%%%%%%%%%%%%%%%%%%%%%%%%%%%%%%%%%%%%%%%%%%%%%%%
%%%%%%%%%%%%%%%%%%%%%%%%%%%%%%%%%%%%%%%%%%%%%%%%%%%%%%%%%%%%%%%%%%%%%%
%%%%%%%%%%%%%%%%%%%%%%%%%%%%%%%%%%%%%%%%%%%%%%%%%%%%%%%%%%%%%%%%%%%%%%
%%%%%%%%%%%%%%%%%%%%%%%%%%%%%%%%%%%%%%%%%%%%%%%%%%%%%%%%%%%%%%%%%%%%%%
\begin{table}
\caption{
Minimum and maximum kinetic energies in fitting 
the transverse mass spectra with a single exponential.
We have tried three sets of parameters, I, II, III and IV.
Each value has the unit in GeV.
}
\label{tab:Mmin}
\begin{tabular}{lrrrrrrrr}
&\multicolumn{2}{c}{I}
&\multicolumn{2}{c}{II}
&\multicolumn{2}{c}{III}
&\multicolumn{2}{c}{IV}
 \\
&$\Delta_{\rm min}$  & $\Delta_{\rm max}$
&$\Delta_{\rm min}$  & $\Delta_{\rm max}$
&$\Delta_{\rm min}$  & $\Delta_{\rm max}$
&$\Delta_{\rm min}$  & $\Delta_{\rm max}$
 \\
\hline
SIS		& 0.0 & 3.0 & 0.5 & 2.0 & 0.5 & 2.0 &  0.5 & 2.0  \\
AGS-RHIC	& 0.0 & 3.0 & 0.5 & 2.0 & 0.8 & 2.0 &  1.0 & 2.0  \\
\end{tabular}
\end{table}

\end{document}